# Observation of the 5p Rydberg states of sulfur difluoride radical by resonance-enhanced multiphoton ionization spectroscopy

**Qun Zhang[a)] and Xiaoguo Zhou**

*Hefei National Laboratory for Physical Sciences at the Microscale and Department of Chemical Physics, University of Science and Technology of China, Hefei, Anhui 230026, P. R. China*

**Quanxin Li, Shuqin Yu, and Xingxiao Ma**

*Department of Chemical Physics, University of Science and Technology of China, Hefei, Anhui 230026, P.R.China*

[a)]Electronic mail: qunzh@ustc.edu.cn

## ABSTRACT

Sulfur difluoride radicals in their ground state have been produced by a "laser-free" pulsed dc discharge of the $SF_6$/Ar gas mixtures in a supersonic molecular beam and detected by mass-selective resonance-enhanced multilphoton ionization (REMPI) spectroscopy in the wavelength range of 408-420 nm. Analyses of the (3 + 1) REMPI excitation spectrum have enabled identification of three hitherto unknown Rydberg states of this radical. Following the Rydberg state labeling in our previous work [see *J. Phys. Chem. A* **102**, 7233 (1998)], these we label the $\tilde{K}(5p_1)$ [$\nu_{0-0}$ = 71 837 cm$^{-1}$, $\omega_1'$ ($a_1$ sym str) = 915 cm$^{-1}$], $\tilde{L}(5p_2)$ [$\nu_{0-0}$ = 72 134 cm$^{-1}$, $\omega_1'$ ($a_1$ sym str) = 912 cm$^{-1}$], and $\tilde{M}(5p_3)$ [$\nu_{0-0}$ = 72 336 cm$^{-1}$, $\omega_1'$ ($a_1$ sym str) = 926 cm$^{-1}$] Rydberg states, respectively. [Origins, relative to the lowest vibrational level of the $\tilde{X}\ ^1A_1$ ground state, and vibrational frequencies of the symmetric S-F stretching mode are suggested by the numbers in brackets.] Photofragmentation process of $SF_2^+ \rightarrow SF^+ + F$ that relates to the REMPI spectrum was discussed.

## I. INTRODUCTION

Resonance-enhanced multilphoton ionization (REMPI) spectroscopy has been widely recognized as a convenient and ultra-sensitive technique for identifying hitherto unknown excited (generally Rydberg) states of (especially non-fluorescent) molecular species, including transient free radicals.[1-3] Such a view gains added credence from the present investigation, in which we report the first observation of the 5$p$ Rydberg states of $SF_2$ radical using this technique.

Sulfur fluoride radicals are believed to play major reactive roles in commercially important plasma processes.[4] It has been known to the semiconductor industry that the silicon etching rates in reactive plasmas formed from $SF_6/O_2$ mixtures are particularly rapid,[5] comparable to or even greater than $CF_4$ plasmas.[6] $SF_2$ radical, as an important reactive intermediate in $SF_6/O_2$ plasmas,[7,8] has received increased attention in the past from both spectroscopists and computational chemists partly because of its potential importance in semiconductor manufacturing.

Although the chemical kinetic and thermodynamic data for $SF_2$ radical remain meager,[9-11] the spectroscopic data for this transient species have been accumulated since its ground state ($\tilde{X}^1A_1$) was identified and characterized by means of microwave spectroscopy[12] and mass spectrometry.[13] The first excited electronic valence state ($\tilde{A}^1B_1$) of $SF_2$ radical was identified by Glinski $et~al.$[14] No Rydberg states were reported until Johnson III and Hudgens directly observed the $\tilde{B}^1B_1(4s)$ and $\tilde{E}(4p)$ Rydberg states by measuring the (2 + 1) and (3 + 1) REMPI spectra, respectively.[15] The irregular three-photon energy interval between the $\tilde{E}(4p)~1_0^2$ and $1_0^3$ bands in their (3 + 1) REMPI spectrum[15] prompted us to have reinvestigated the $\tilde{E}$ (4*p*) Rydberg state. Our (2 + 1) REMPI spectrum associated with the $\tilde{E}$ (4*p*) Rydberg state revealed that another $\tilde{F}$ (4*p*) Rydberg state should reside very close to the $\tilde{E}$ (4*p*) Rydberg state, and the assignments for the two 4*p* Rydberg states naturally solved the irregular band interval puzzle.[16] In addition, four higher Rydberg states, $\tilde{G}$ (3*d*), $\tilde{H}$ (3*d*), $\tilde{I}$ (3*d*), and $\tilde{J}\,^1B_1$ (5*s*), have also been identified from our (2 + 1) REMPI spectra.[16] The valence $\tilde{C}$ state of predissociative character was first observed by Johnson III and Hudgens,[15] and later characterized by us.[17,18] Recent large-scale multireference single and double excitation configuration interaction (MRSDCI) calculations by Liu *et al.*,[19] in comparison with experimental observations,[15,16] indicted that the $\tilde{C}$, $\tilde{E}$ (4*p*), $\tilde{F}$ (4*p*), $\tilde{G}$ (3*d*), $\tilde{H}$ (3*d*), and $\tilde{I}$ (3*d*) states of undetermined state symmetries may probably be assigned to $3\,^1B_1$, $2\,^1A_2$, $2\,^1A_1$, $3\,^1A_1$, $4\,^1B_1$, and $1\,^1B_2$, respectively. Besides the two known $\tilde{A}\,^1B_1$ and $\tilde{C}$ valence states, a new excited valence state, $\tilde{B}'$, has also been observed to reside between the $\tilde{A}\,^1B_1$ and $\tilde{B}\,^1B_1$ (4*s*) states.[20]

The electronic band systems of $SF_2$ radical have hitherto been well established up to the $\tilde{J}\,^1B_1$ (5*s*) Rydberg state.[16] It comes naturally to ask if other higher Rydberg states of $SF_2$ radical can also be observable using the sensitive REMPI technique.

The (2 + 1) REMPI spectrum of $SF_2$ radical between the laser wavelength range of 263-275 nm (corresponding to a two-photon energy region of 72 727-76 046 cm$^{-1}$) we observed in our previous work (see Figure 7 in Ref. 16) shows a broad diffuse structure, from which one can hardly resolve the expected band systems associated with the 4*d*, 4*f*, 6*s*, 6*p*, and/or even higher Rydberg states. Slightly lower than this energy region should lie the 5p Rydberg state(s). Considering that the members belonging to a Rydberg series of the same type usually exhibit absorption intensities in proportion to $1/n^2$ (n, being the principle quantum number of the excited electron),[1] and that the well resolved REMPI spectra attributable to the 4p Rydberg states of $SF_2$ radical have been successfully detected,[15,16] we expect that the resolved REMPI spectra associated with the 5p Rydberg state(s) may also be obtainable. Unfortunately, we failed to see spectral features associated with the 5p Rydberg state(s) by recording the (2 + 1) REMPI spectrum in the expected wavelength range of 272-280 nm (corresponding to a two-photon energy region of 71 429-73 529 cm$^{-1}$). It is worth contemplating: (1) what caused our failure using a (2 + 1) REMPI scheme, and (2) why the 5p Rydberg states have not been reported in the previous work by Johnson III and Hudgens using a (3 + 1) REMPI scheme,[15] for it is obviously not a hard task to extend their used wavelength range of 450-490 nm (where they observed the $\tilde{E}(4p)$ Rydberg state) down to a range of wavelengths shorter than 420 nm (where the 5p Rydberg state(s) are expected to reside).

In this paper we present the first spectroscopic observation of the 5p Rydberg states of $SF_2$ radical using a (3 + 1) REMPI excitation scheme as well as a reasonable explanation for under what conditions the REMPI spectrum of interest can be obtained.

## II. EXPERIMENT

The experimental apparatus and procedure has been described in detail elsewhere.[16] $SF_2$ radicals were produced by the pulsed dc discharge in a pulsed molecular beam of $SF_6$ (30% in Ar). To obtain a stable discharge in the pulsed beam, a special design based on some laser discharge modes (such as those adopted in commercially available excimer and TEA $CO_2$ lasers) was utilized. The discharge was initiated between four pairs of closely spaced tungsten needles mounted on between a pair of parallel copper electrode plates, and a stable glow discharge was then created between the plates. The background ions produced during the discharge were effectively eliminated by a set of ion reflectors. The neutral radical products were allowed to supersonically expand into the source region of a time-of-flight (TOF) mass spectrometer where they were interrogated by the output of a Nd:YAG pumped dye laser (~2 mJ/pulse) which was focused with a 170-mm lens. The dye laser wavelength was calibrated against known neon and argon atomic transitions using optogalvanic spectroscopy.[21,22] Ions produced by the REMPI process were detected using a microchannel plate (MCP) detector situated at the end of the TOF mass spectrometer. The REMPI spectra of $SF_2$ radical were recorded by monitoring the portion of the ion current appearing in the *m/z* 70 mass channel as a function of the excitation wavelength.

## III. RESULTS

Figure 1 exhibits the typical TOF mass spectra recorded under the following conditions: (a) only the dc discharge was on, (b) only the 417.61 nm laser irradiation was on, and (c) both the dc discharge and the 417.61 nm laser irradiation were on. The *m/z* 70 mass channel corresponding to the $^{32}SF_2^+$ ions appeared sorely in the case (c), which indicated that the mass channel of interest should be contributed to neither the residual background ions produced during the dc discharge nor the MPI products of the parent $SF_6$ molecules. In order to further identify the spectral carrier, we have also recorded a REMPI spectrum in the wavelength range of 325-365 nm under the same discharge conditions. The well resolved vibrational progressions in thus obtained (2 + 1) REMPI spectrum turned out to be associated with the $\tilde{B}\,^1B_1$ and $\tilde{C}$ states of the $SF_2$ radical, which is in excellent accord with our previous observations[16] as well as the earlier report by Johnson III and Hudgens.[15] This further evidenced that the ion signals in the *m/z* 70 mass channel we recorded should arise unambiguously from the neutral $^{32}SF_2$ radicals.

Figure 2 shows the *m/z* 70 (3 + 1) REMPI excitation spectrum of $SF_2$ radical in the laser wavelength range of 408-420 nm. We have labeled the electronic states that yield this spectrum the $\tilde{K}$, $\tilde{L}$, and $\tilde{M}$ states, respectively, following our previous labeling.[16] Table I lists the band maxima observed between 408 and 420 nm, three-photon energies, band intervals of each progression, and assignments of the three states. The assignments characterize the three states with the spectroscopic values: $\nu_{0-0}$ = 71 837 cm$^{-1}$ and $\omega'_1$ ($a_1$ sym str) = 915 cm$^{-1}$ for the $\tilde{K}$ state; $\nu_{0-0}$ = 72 134 cm$^{-1}$ and $\omega'_1$ ($a_1$ sym str) = 912 cm$^{-1}$ for the $\tilde{L}$ state; $\nu_{0-0}$ = 72 336 cm$^{-1}$ and $\omega'_1$ ($a_1$ sym str) = 926 cm$^{-1}$ for the $\tilde{M}$ state. [$\nu_{0-0}$ denotes the band origins, $\omega'_1$ ($a_1$ sym str) the vibrational frequencies of symmetric S-F stretching mode.] In an attempt to find more vibrational bands for each state and hence better characterize these states, we have extended the used wavelength range to below 408 nm. Unfortunately, it turned out that no more resolved spectral features were observed (for explanation, see the next section).

If we persist with the adiabatic ionization potential ($IP_a$) value of $SF_2$ (10.08 eV) reported previously,[23] between 408 and 420 nm $SF_2$ must absorb four photons to ionize and form the $\tilde{X}\,^2B_1$ ground state cation. The data indicate that the resonant states are prepared by the simultaneous absorption of three photons. [The three-photon resonance was further justified by a photon index measurement, *i.e.*, measuring the *m/z* 70 ion signal intensity as a function of the laser power.] Assuming three-photon preparation, the $\omega_1'$ ($a_1$ sym str) frequencies (~920 cm$^{-1}$) of the $\tilde{K}$, $\tilde{L}$, and $\tilde{M}$ states we derived are quite similar to the vibrational frequency of the $\tilde{X}\,^2B_1$ ground state cation, $\omega_1$ = 935 (40) cm$^{-1}$.[23] The close similarity implies that each of the three newly observed states may arise as a result of promoting an electron from the highest occupied molecular orbital (HOMO) in the $\tilde{X}\,^1A_1$ ground state of SF$_2$ radical to an orbital which is largely nonbonding. Given the energies involved, the excited orbitals are likely to be predominantly Rydberg in character. The type of Rydberg orbitals involved in the $\tilde{K}$, $\tilde{L}$, and $\tilde{M}$ states can be determined by calculating a quantum defect for each origin transition using the reduced Rydberg formula, $h\upsilon_{0-0} = \text{IP}_a - R/(n-\delta)^2$, where $\nu_{0-0}$ denotes the band origin, IP$_a$ the adiabatic ionization potential (81 301 cm$^{-1}$ [23] for SF$_2$), $R$ the Rydberg constant (109 737 cm$^{-1}$), $n$ the principle quantum number, and $\delta$ the quantum defect value. The only reasonable solution gives $n$ = 5, and $\delta$ = 1.59, 1.54, and 1.50 for the $\tilde{K}$, $\tilde{L}$, and $\tilde{M}$ states, respectively. Because the quantum defects for *ns*, *np*, *nd*, and *nf* Rydberg states centered on sulfur atom should lie near 2.0, 1.6, 0.08, and 0.06,[24] the solution suggests that the three newly observed states should be assignable in terms of excitation to a 5*p* Rydberg orbital, *i.e.*, they should all correspond to a 5*p*-type Rydberg state with the same $\tilde{X}\,^2B_1$ cation core.

The $\tilde{X}\,^1A_1$ ground state of SF$_2$ radical is described in C$_{2v}$ symmetry group by the electronic configuration:

$$1a_1^2 2a_1^2 1b_2^2 3a_1^2 1b_1^2 4a_1^2 2b_2^2 5a_1^2 3b_2^2 6a_1^2 2b_1^2 7a_1^2 4b_2^2 1a_2^2 5b_2^2 8a_1^2 3b_1^2.$$

The 5p Rydberg orbitals are of symmetry $b_1(5p_x)$, $b_2(5p_y)$, and $a_1(5p_z)$, which give rise to state symmetries of $^1A_1$, $^1A_2$, and $^1B_1$, respectively. The spectral features arise from a three-photon excitation:

$$...5b_2^2 8a_1^2 3b_1^2 \; SF_2 \; (\tilde{X}\,^1A_1) + 3h\upsilon \rightarrow ...5b_2^2 8a_1^2 3b_1^1 5pb_1^1 \; SF_2 \; (^1A_1),$$
$$...5b_2^2 8a_1^2 3b_1^2 \; SF_2 \; (\tilde{X}\,^1A_1) + 3h\upsilon \rightarrow ...5b_2^2 8a_1^2 3b_1^1 5pb_2^1 \; SF_2 \; (^1A_2), \text{ and}$$
$$...5b_2^2 8a_1^2 3b_1^2 \; SF_2 \; (\tilde{X}\,^1A_1) + 3h\upsilon \rightarrow ...5b_2^2 8a_1^2 3b_1^1 5pa_1^1 \; SF_2 \; (^1B_1),$$

followed by an ionization step through the processes:

$$...5b_2^2 8a_1^2 3b_1^1 5pb_1^1 \; SF_2 \; (^1A_1) \xrightarrow{-e^-} ...5b_2^2 8a_1^2 3b_1^1 \; SF_2^+ \; (\tilde{X}\,^2B_1),$$
$$...5b_2^2 8a_1^2 3b_1^1 5pb_2^1 \; SF_2 \; (^1A_2) \xrightarrow{-e^-} ...5b_2^2 8a_1^2 3b_1^1 \; SF_2^+ \; (\tilde{X}\,^2B_1), \text{ and}$$
$$...5b_2^2 8a_1^2 3b_1^1 5pa_1^1 \; SF_2 \; (^1B_1) \xrightarrow{-e^-} ...5b_2^2 8a_1^2 3b_1^1 \; SF_2^+ \; (\tilde{X}\,^2B_1).$$

Since the ionization step requires only ~9 700 cm$^{-1}$ (energy difference between IP$_a$ and three photons of 419 nm), SF$_2$ in the 5p Rydberg states can ionize after absorbing one laser photon, i.e., a (3 + 1) REMPI mechanism accounts for the spectrum shown in Fig. 2. However, the spectrum we recorded does not reveal the symmetry of the upper states, thus we tentatively label them the $\tilde{K}(5p_1)$, $\tilde{L}(5p_2)$, and $\tilde{M}(5p_3)$ states in order of increased state energies, as shown in Fig. 2 and Table I.

## IV. DISCUSSION

It is worth noting that discernable ion signals appeared in the m/z 51 mass channel which corresponds to $^{32}$SF$^+$ ions when the m/z 70 (3 + 1) REPMI spectrum was recorded in the wavelength range of 408-414.6 nm (corresponding to the portion between 72 359 and 73 529 cm$^{-1}$ in Fig. 2). The *m/z* 51 "daughter" REMPI spectrum in this wavelength range turned out to carry features attributable to $^{32}SF_2$, which implies that in the corresponding four-photon energy region of 96 479-98 039 cm$^{-1}$ (relative to the lowest vibrational level of the $\tilde{X}\,^1A_1$ ground state of $SF_2$, or 15 178-16 738 cm$^{-1}$ relative to the lowest vibrational level of the $\tilde{X}\,^2B_1$ ground state of $SF_2^+$) may occur photofragmentation of $SF_2^+$ cations, *i.e.*, a process of resonance-enhanced multiphoton ionization followed by dissociation (REMPID) leading to $SF_2^+$ ($m/z$ 70) $\rightarrow$ $SF^+$ ($m/z$ 51) + F. In another experiment we have conducted with an aim to reinvestigate the 4*p* Rydberg states of $SF_2$ radical using a (2 + 1) REMPI mechanism over the wavelength range of 293-323 nm, the similar phenomenon was observed in the wavelength range of 293-311.1 nm.[25] The onset wavelength (311.1 nm) of the photofragmentation of $SF_2^+$ cations corresponds to a three-photon energy of 96 432 cm$^{-1}$, which conforms quite nicely to the present observation (a four-photon energy of 96 479 cm$^{-1}$), *i.e.*, the two REMPID ($SF_2$ $\rightarrow$ $SF_2^+$ $\rightarrow$ $SF^+$ + F) processes under different excitation schemes [*i.e.*, (3 + 1) *vs.* (2 + 1)] commence at about the same energy position which lies between the ground $\tilde{X}\,^2B_1$ state and the first excited $\tilde{A}\,^2A_1$ state of $SF_2^+$ cation.[23,26] It seems farfetched to explain this coincidence as only an accidental event. It may indicate that the above ionization threshold continua reached by both REMPI schemes provide preferential treatment to the opening of the $SF_2^+$ $\rightarrow$ $SF^+$ + F photofragmentation channel. However, we cannot rule out the possibility of one more photon absorption for both schemes [ *i.e.*, (3 + 2) *vs.* (2 + 2) instead of (3 + 1) *vs.* (2 + 1)]. When compared with its neutral counterpart, the $SF_2^+$ cation is found still far from being well characterized.[26] *Ab initio* calculations in an attempt to reveal the possible mechanisms behind the phenomena aforementioned are being carried out in our group.[25]

In addition, the tendency of the REMPID process was noticed to become heavier at higher laser intensities, *i.e.*, the $SF_2^+$ "parent" ion signals decrease while the $SF^+$ "daughter" ion signals increase drastically with increased photon flux. The laser intensity used was ~2 mJ/pulse for both experiments, the focal length $f$ of the lenses used, however, was quite different: $f = 300$ mm for the (2 + 1) experiment, while $f = 170$ mm for the (3 + 1) experiment. The spectral features shown in Fig. 2 vanished when a lens of $f = 300$ mm was used, which can be explicitly accounted for the fact that the simultaneous three-photon absorption rate is much slower than the simultaneous two-photon absorption rate,[1] *i.e.*, in order to detect electronic states using a (3 + 1) REMPI excitation scheme one must ensure a sufficiently intense laser. On the other hand, application of a tighter focusing with a lens of $f$ shorter than 100 mm resulted in strong photofragmentation of $SF_2^+$ cations, which in turn precluded a direct measurement of the *m/z* 70 "parent" REMPI spectrum. A moderate focusing condition ($f = 170$ mm) deliberately chosen in the present study yielded the *m/z* 70 REMPI spectrum of $SF_2$ radical shown in Fig. 2.

To achieve an appropriate compromise between suppressing the detrimental REMPID process and ensuring sufficient photon flux plays a critical role in obtaining the spectrum reported here, ignorance or incomplete investigation of which may partly explain why the 5*p* Rydberg states have not been reported in the previous work by Johnson III and Hudgens using the same (3 + 1) REMPI scheme.[15] Note that they claimed no evidence for photofragmentation of $SF_2^+$ cations was observed when their (3 + 1) REMPI spectrum assignable to the $\tilde{E}(4p)$ Rydberg state was recorded in the wavelength range of 450-490 nm.[15] Their observations do not contradict ours, in that the onset four-photon energy position (~96 479 cm$^{-1}$) of the $SF_2^+ \rightarrow SF^+ + F$ photofragmentation we observed in this work lies much higher than the four-photon energy region between 81 633 and 88 889 cm$^{-1}$ which corresponds to their 450-490 nm laser wavelength range.

Furthermore, we have ever expected that, provided the (2 + 1) REMPI scheme works the spectral features associated with the 5p Rydberg state(s) of $SF_2$ radical may also be observable in the wavelength range of 272-280 nm, considering that the estimated origin of the 5p Rydberg state(s) lies near 278.4 nm in terms of a two-photon preparation. This expectation came naturally from the fact that we have observed well resolved spectra assignable to the two-photon resonant transitions of $\tilde{E}(4p) \leftarrow\leftarrow \tilde{X}^1A_1$ and $\tilde{F}(4p) \leftarrow\leftarrow \tilde{X}^1A_1$ using a (2 + 1) REMPI scheme in the wavelength range of 295-325 nm.[16] However, a careful survey over the wavelength range of 272-280 nm under a variety of focusing conditions failed to make us resolve spectral features arising from the desirable 5p Rydberg state(s). It may implicitly be accounted for the fact that the undesirable REMPID process involved in this (2 + 1) experiment turned out much harder to be suppressed without losing considerable *m/z* 70 "parent" ion signals than that in the present (3 + 1) experiment.

In terms of three-photon absorption, the REMPID process in the present study was found to show up from 72 359 cm$^{-1}$ (corresponding to three 414.6 nm photons, right between three $0_0^0$ bands and three $1_0^1$ bands, see Fig. 2). When the laser wavelengths were scanned to below 408 nm, it turned out even harder to suppress the photofragmentation by only varying the laser intensity, thereby preventing us from obtaining reliable $1_0^n$ ($n \geq 2$) bands (and/or bands related to $\omega_2'$). Owing to the limited number of vibrational bands observed in the present REMPI spectrum, the vibrational frequencies of $\omega_1'$ ($a_1$ sym str) for each state reported here can only be suggested. To obtain more accurate and comprehensive spectroscopic constants for these newly identified Rydberg states of $SF_2$ radical requires further experimental and theoretical investigations.

Last but not least, it is conceivable that an indirect measurement of the "daughter" REMPI spectra may reveal the same spectral features as the direct measurement of the "parent" REMPI spectra does. This is in principle the case, given a REMPID instead of a REMPDI (resonance-enhanced multiphoton dissociation followed by ionization) [17] process is involved. In practice, however, it turned out from our experimental observations that the "daughter" REMPI spectra usually possess a worse signal-to-noise ratio than the "parent" REMPI spectra, which in turn makes their resolution deteriorated. More importantly, the "daughter" REMPI spectra may suffer (quite commonly) a detrimental loss of useful spectroscopic information carried by the neutral "parent" species, as in the cases discussed above.

## V. CONCLUSION

A "laser-free" pulsed dc discharge technique was utilized to generate the neutral sulfur difluoride radicals in a supersonic molecular beam. The (3 + 1) REMPI spectrum of $SF_2$ radical was observed in the excitation wavelength range of 408-420 nm. Three 5$p$ Rydberg states of $SF_2$ radical were identified and characterized with suggested spectroscopic constants. Insights into the REMPID process afflicting the acquisition of a well resolved $SF_2$ "parent" REMPI spectrum were put in detail, which may facilitate a further exploration of other hitherto unknown (*e.g.*, 4*d*, 4*f*, 6*s*, and 6*p*) Rydberg states of this particular radical, and may as well be instructive for revealing new high-lying electronic states of other molecular species by means of REMPI spectroscopy.

## ACKNOWLEDGEMENTS

This research was made possible by a financial support from the National Natural Science Foundation of China.  Q. Z. would like to thank the University of Science and Technology of China for awarding a start-up fund for the returned overseas Chinese scholars.  The authors have enjoyed useful discussions with Professor Y. Chen and Professor S.-X. Tian.  Experimental assistance from Professor J.-N. Shu is also gratefully acknowledged.

## **REFERENCES**

[1]J. W. Hudgens, in *Advances in Multiphoton Processes and Spectroscopy*, edited by S. H. Lin (World Scientific, Singapore, 1988), Vol. 4, p. 171.

[2]M. N. R. Ashfold, S. G. Clement, J. D. Howe, and C. M. Western, J. Chem. Soc. Faraday Trans. **89**, 1153 (1993).

[3]M. N. R. Ashfold and J. D. Howe, Annu. Rev. Phys. Chem. **45**, 57 (1994).

[4]D. M. Manos and D. L. Flamm, *Plasma Etching: An Introduction* (Academic, Boston, 1989).

[5]R.  d′ Agostino and D. L. Flamm, J. Appl. Phys. **52**, 162 (1981).

[6]S. Park, C. Sun, and R. J. Purtell, J. Vac. Sci. Technol. B **5**, 1372 (1987).

[7]G. D. Griffin, C. E. Easterly, I. Sauers, H. W. Ellis, and L. G. Christophorou, Toxicol. Environ. Chem. **9**, 139 (1984).

[8]K. R. Ryan and I. C. Plumb, Plasma Chem. Plasma Proc. **8**, 263 (1988).

[9]J. T. Herron, J. Phys. Chem. Ref. Data **16**, 1 (1987).

[10]R. J. Van Brunt and J. T. Herron, IEEE Trans. Electrical Insul. **25**, 75 (1990).

[11]K. K. Irikura, J. Chem. Phys. **102**, 5357 (1995).

[12]D. R. Johnson and F. X. Powell, Science **164**, 950 (1969).

[13]F. Seel, E. Heinrich, W. Gombler, and R. Budenz, Chimia **23**, 73 (1969).

[14]R. J. Glinski, Chem. Phys. Lett. **129**, 342 (1986); R. J. Glinski and C. D. Taylor, Chem. Phys. Lett. **155**, 511 (1989); R. J. Glinski, C. D. Taylor, and F. W. Kutzler, J. Phys. Chem. **94**, 6196 (1990).

[15]R. D. Johnson III and J.W. Hudgens, J. Phys. Chem. **94**, 3273 (1990).

[16]Q. X. Li, J. N. Shu, Q. Zhang, S. Q. Yu, L. M. Zhang, C. X. Chen, and X. X. Ma, J. Phys. Chem. A **102**, 7233 (1998).

[17]X. G. Zhou, Q. X. Li, Q. Zhang, S. Q. Yu, Y. Su, C. X. Chen, and X. X. Ma, J. Elec. Spec. Relat. Phenom. **108**, 135 (2000).

[18]X. G. Zhou, S. Q. Yu, Z. Y. Sheng, X. X. Ma, Y. J. Liu, M. B. Huang, and Q. X. Li, Surf. Rev. Lett. **9**, 69 (2002).

[19]Y. J. Liu, M. B. Huang, X. G. Zhou, and S. Q. Yu, Chem. Phys. Lett. **345**, 505 (2001).

[20]Q. X. Li, Q. Zhang, J. N. Shu, S. Q. Yu, Q. H. Song, C. X. Chen, X. X. Ma, Chem. Phys. Lett. **305**, 79 (1999).

[21]A. R. Striganov and N. S. Sventitskii, *Tables of Spectral Lines of Neutral and Ionized Atoms* (Plenum, New York, 1968).

[22]O. S. King, P. K. Schenck, and J. C. Travis, Appl. Opt. **16**, 2617 (1977).

[23]D. M. De Leeuw, R. Mooyman, and C. A. De Lange, Chem. Phys. **34**, 287 (1978).

[24]S. T. Manson, Phys. Rev. **182**, 97 (1969).

[25] Q. Zhang, S. X. Tian, *et al.*, "On the photofragmentation of $SF_2$ cation: Experiment and *ab initio* calculations," J. Chem. Phys. (to be submitted).

[26] E. P. F. Lee, D. K. W. Mok, F. T. Chau, and J. M. Dyke, J. Chem. Phys. **125**, 104304 (2006).

# TABLES

TABLE I. Band maxima, assignments, and spacings observed in the (3 + 1) REMPI spectrum of $SF_2$ radical between 408 and 420 nm. [$0_0^0$ denotes the origin ($v_i' = 0 \leftarrow\leftarrow\leftarrow v_i'' = 0$) ($i$ = 1, 2, 3) band, and $1_0^1$ the ($v_1' = 1 \leftarrow\leftarrow\leftarrow v_1'' = 0$, and $v_i' = 0 \leftarrow\leftarrow\leftarrow v_i'' = 0$) ($i$ = 2, 3) band for each state.]

| Assignment | Band Max ($\lambda_{air}$), nm | State energy cm$^{-1}$ | Energy rel. to each state origin $0_0^0$, cm$^{-1}$ |
|---|---|---|---|
| $\tilde{K}(5p_1)$ $0_0^0$ | 417.61 | 71 837 | 0 |
| $\tilde{K}(5p_1)$ $1_0^1$ | 412.36 | 72 752 | 915 |
| $\tilde{L}(5p_2)$ $0_0^0$ | 415.89 | 72 134 | 0 |
| $\tilde{L}(5p_2)$ $1_0^1$ | 410.70 | 73 046 | 912 |
| $\tilde{M}(5p_3)$ $0_0^0$ | 414.73 | 72 336 | 0 |
| $\tilde{M}(5p_3)$ $1_0^1$ | 409.49 | 73 262 | 926 |

# **FIGURES**

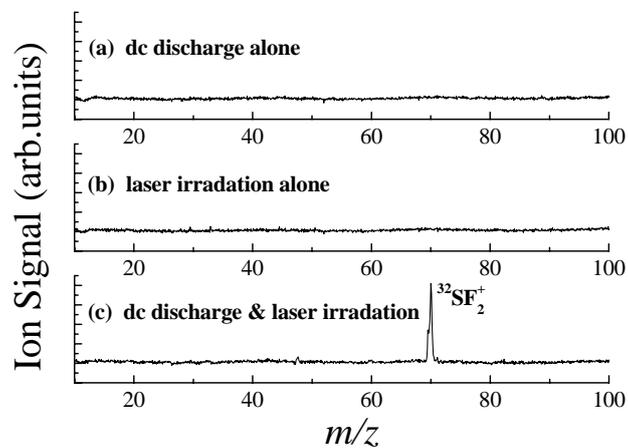

**FIG. 1**. Typical TOF mass spectra under three cases: (a) the dc discharge alone, (b) the 417.61 nm laser irradiation alone, and (c) the dc discharge together with the 417.61 nm laser irradiation. The $^{32}SF_2^+$ (*m/z* 70) ion signals appeared only in the case (c).

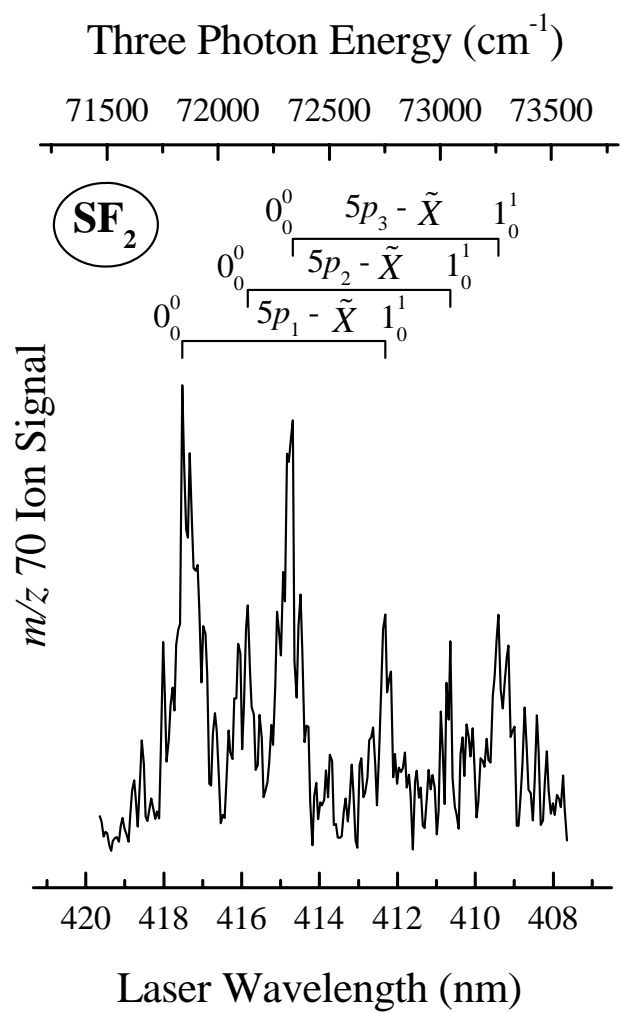

**FIG. 2.** (3 + 1) REMPI excitation spectrum of the $^{32}SF_2$ radical (*m/z* 70) between 408 and 420 nm.